\documentstyle[aas2pp4]{article}
\newcommand{\grad}{\mathop{\rm grad}\nolimits}
\def\dfrac#1#2{{\displaystyle#1\over\displaystyle#2}}

\begin{document}

\title{
Impact of a Binary System Common Envelope on Mass Transfer through
the Inner Lagrange Point}

\author{D.V.Bisikalo\altaffilmark{1}}
\affil{Institute of Astronomy of the Russian Acad. of Sci.,
Moscow, Russia}

\author{A.A.Boyarchuk}
\affil{Institute of Astronomy of the Russian Acad. of Sci.,
Moscow, Russia}

\author{O.A.Kuznetsov\altaffilmark{2}}
\affil{Keldysh Institute of Applied Mathematics, Moscow, Russia}

\author{V.M.Chechetkin}
\affil{Keldysh Institute of Applied Mathematics, Moscow, Russia}

\altaffiltext{1}{\large E-mail address: {\it bisikalo@inasan.rssi.ru}}
\altaffiltext{2}{\large E-mail address: {\it kuznecov@spp.keldysh.ru}}

\begin{abstract}
Results of numerical simulations of the impact of a common
envelope on the matter flow pattern near the outflowing
component in a semidetached binary system are presented.
Three-dimensional modeling of the matter transfer gas dynamics
in a low-mass X-ray binary X1822--371 enable
investigation of the structure of flows in the vicinity of the
inner Lagrange point $L_1$. Taking into account the common
envelope of the system substantially changes the flow pattern
near the Roche surface of the outflowing component. In a
stationary regime, accretion of common envelope gas is observed
over a significant fraction of the donor star's surface, which
inhibits the flow of gas along the Roche surface to $L_1$. The
change in the flow pattern is particularly significant near
$L_1$, where the stream of common envelope gas strips matter off
the stellar surface.  This, in turn, significantly increases (by
an order of magnitude) the gas flow from the donor surface in
comparison with the estimates of standard models.

\end{abstract}

\section{Introduction}

   The observational manifestations of interaction in
semidetached binary systems (cataclysmic binaries, low-mass
X-ray binaries, and supersoft X-ray sources) are extremely
interesting.  Comparisons of computational results from
numerical simulations of matter flows in these systems with
observational data provide a basis for studying and
understanding the physical processes occurring in these binaries.
As a rule, numerical simulations of semidetached systems have
been performed using the generally well-justified assumptions
that the orbits of the system components are circular and the
components' rotation is synchronous with their orbital motion.
In this case, the standard scenario for mass transfer in the
system is as follows. In the course of its evolution, the donor
star fills its critical surface (which, under the outlined
assumptions, coincides with the Roche surface in the restricted
three-body problem), and mass begins to flow through the
vicinity of the inner Lagrange point $L_1$, where the pressure
gradient is not balanced by the gravitation force.

   The flow pattern in semidetached systems has been
investigated using both analytical [1--3] and numerical [4--15]
models.  However, in all these studies, the impact of the common
envelope of the system on the structure of gaseous flows was
either not taken into account at all, or was taken into account
not entirely correctly. The morphology of flows in binary
systems allowing for the presence of a common envelope was first
considered by us in [16] (henceforth, Paper I). There, as well
as in this paper, the "common envelope" of the system refers to
the gas filling the space between the two components of the
system and not involved in the accretion process (i.e., not
belonging to the accretion disc). Our results of
three-dimensional numerical simulations of a low mass X-ray
binary X1822--371 in Paper I showed the contribution
of the common envelope to the formation of flow structures in
the system to be significant.

  The presence of a common envelope also influences the flow
pattern in the vicinity of the donor star, which, in turn, is
reflected in the mass exchange rate in the system. This problem
is extremely important, because though the observational
manifestations of interaction primarily depend on the general
flow pattern studied in Paper I, the evolution of the system is
determined by the mass exchange parameters. Here, we present the
results of an analysis based on the calculations described in
Paper I, but with the main accent on a detailed consideration of
the influence of the common envelope on the structure of gas
flows near the Roche surface of the outflowing component.

\section{Flow Structure in the Vicinity of $L_1$.\\ Results of
Standard Models}

  Analytical studies of the mass transfer process in
semidetached systems make it possible to draw conclusions
concerning the parameters of flows of matter (streams) in such
systems. The first detailed analysis of matter flows in the
vicinity of $L_1$ was carried out by Lubow and Shu [1], who
estimated the basic characteristics of the stream using a
perturbation method.  In later papers, the stream parameters
were refined using Bernoulli integral analysis, and the
dependence of the mass transfer rate on the degree of overflow
of the Roche lobe by the donor star was found [2,3].  The basic
characteristics of the streams obtained in various studies
(including numerical calculations, e.g., [17]) differ only in
details, and are currently widely used as standard results in
analyses of mass transfer in semidetached binary systems (see,
e.g., the reviews [18, 19]). Unfortunately, none of these
studies took into account the impact of the common envelope of
the system on the flow pattern near the donor star. The aim of
our study is to consider the contribution of the common envelope
and to obtain new stream parameter estimates using the results
of three-dimensional numerical simulations of the flow in a
low-mass X-ray binary system (these will be presented in Section
3). However, when analyzing changes in the flow pattern
associated with the presence of a common envelope, it is
expedient to first consider the generally accepted values for
flow parameters in semidetached systems.

   We will first define the reference frame for our analysis. We
place the origin of the coordinate frame at the donor star
center of mass. In our Cartesian coordinate system $(x,y,z)$, we
will direct the $X$ axis along the line connecting the centers
of the stars, the $Z$ axis along the rotation axis, and the $Y$
axis so that the resulting system is right-handed. We will set
the characteristic scale of the coordinate system by placing the
secondary star at the point $(A,0,0)$, where $A$ is the distance
between the two components.  This means that the center of mass
of the system is at the point $(\mu A,0,0)$, where
$\mu=M_2/(M_1+M_2)$ is the ratio of the accretor's mass to the
total mass of the system.

   It is assumed in standard analytical models that the matter
at the surface of the donor star (which coincides with the Roche
lobe) is characterized by some temperature (or local sound speed
$c_0$) and density. The kinetic energy of this matter
($\sim c^2_0$) determines the degree of overflow of the Roche
lobe.  The local sound speed of the gas for typical semidetached
binaries is much less than the orbital velocity of the system
components $v_{orb}\sim \Omega A$, and their ratio is,
as a rule, used as a small parameter in analyses. Proceeding
from consideration of the energetic conditions of the gas at the
Roche surface, it is not difficult to estimate the
characteristic dimensions of the matter stream that forms in the
vicinity of $L_1$. The store of kinetic energy of the gas allows
it to move in the $YZ$ plane passing through $L_1$. Equating the
potential energy difference and the specific kinetic energy, we
can derive an equation for the shape of the stream in the
vicinity of $L_1$, similar to the corresponding equation from
[1]. Following the adopted scheme for determining the $YZ$ cross
section of the stream passing through $L_1$, we can write:

$$
\Phi - {\Phi}_{(x_{L_1},0,0)} = c_0^2,\eqno(1)
$$
where $\Phi$ is the force field potential in a reference frame
rotating with the angular velocity $\Omega$ of the system's
rotation about the common center of mass, in the absence of
other external forces, excluding gravity. The potential in the
Roche approximation has the form:

$$
\begin{array}{ccc}
\Phi(x,y,z)&=&-\frac{\displaystyle
GM_1}{\displaystyle\sqrt{x^2+y^2+z^2}}\\
~\\
&&-\frac{\displaystyle
GM_2}{\displaystyle\sqrt{(x-A)^2+y^2+z^2}}\\
~\\
&&-\frac{\displaystyle\Omega^2}{\displaystyle 2}
\left((x-\mu A)^2+y^2\right),
\end{array}
$$
where the term $-\frac{\displaystyle\Omega^2}{\displaystyle 2}\left((x-\mu
A)^2+y^2\right)$ is associated with the centrifugal force.

Expanding expression (1) into a Taylor series in $y$ and $z$ in
the vicinity of $L_1$ and using the condition

$$
\left.\grad~\Phi\right|_{(x_{L_1},0,0)}=0,
$$
it is not difficult to show that the stream cross section is an ellipse with
semiaxes

$$
a=\sqrt{g_y(\mu)}\frac{c_0}{\Omega}\sim\frac{c_0}{\Omega}
\eqno(2)
$$
and

$$
b=\sqrt{g_z(\mu)}\frac{c_0}{\Omega}\sim\frac{c_0}{\Omega},
\eqno(3)
$$
where $g_y(\mu)\sim  1$  and  $g_z(\mu)\sim 1$ are the
coefficients of the expansion of the Roche potential in the
Taylor series in $y$ and $z$ in the vicinity of $L_1$, determined
by the relation:

$$
\Phi=\Phi_{L_1}+g_y(\mu)\Omega^2 y^2+ g_z(\mu)\Omega^2 z^2.
$$

  Apart from considering the stream dimensions at the point
$L_1$, it is also of interest to determine its deflection as a
result of the Coriolis force.  According to the analysis of
Lubov and Shu [1], the stream deflection angle is

$$
\cos(2\theta_s)=-\frac{4}{3g_\theta}+\left(1-\frac{8}{9g_\theta}\right
)^{1/2},
$$
where

$$
g_\theta(\mu)=\frac{\mu}{(x_{L_1}/A)^3}+\frac{1-\mu}{(1-x_{L_1}/A)^3}.
$$
The rotation angle of the stream depends on the mass ratio of
the components, and, in the simplest case (without taking into
account the stream's interaction with the common envelope gas),
it lies in the range from $-28.37^\circ$ to $-19.55^\circ$.

   In more thorough analytical models (see, e.g. [2,3]), the
Bernoulli integral for the matter leaving the donor star surface
is used to determine the stream parameters. The conservation of
this integral along a stream line,

$$
\Phi+\frac{u^2}{2}+\frac{c^2}{\gamma-1}=
\Phi_{L_1}+\frac{u_0^2}{2}+\frac{c_0^2}{\gamma-1},
$$
and the assumption that the gas flow velocity is equal to the
sound speed (the Mach number is 1) can be used to derive both
the stream dimensions and the cross-sectional distribution of
parameters across the stream. The stream dimensions obtained in
this approximation differ from (2) and (3) by a factor of
$\sqrt{\frac{1}{\gamma-1}}$, however, analysis of the density
distribution across the stream shows that the stream profile
consists of a dense core whose characteristic dimensions are
close to expressions (2) and (3) and a rarefied peripheral
region.

   The dimensions of the stream are of fundamental importance
for determining the characteristics of the stream of matter
flowing from the vicinity of $L_1$.  The values for the total
mass flow from the system in various approaches are essentially
equal (within a factor of two), and can be determined from the
relation

$$
\dot M = \rho_0 c_0 S, \eqno(4)
$$
where $\rho_0$ and $c_0$ are the gas density and sound speed at
the surface of the outflowing star and

$$
S=\pi a b = \pi\sqrt{g_y(\mu)g_z(\mu)}\frac{c_0^2}{\Omega^2}
\sim \frac{c_0^2}{\Omega^2}.
$$
is the area of the stream cross section. Since for a polytropic
gas $\rho=\left(\dfrac{c_0^2}{\gamma K}\right)
^{\frac{1}{\gamma-1}}$, the final expression for the dependence
of the mass flow on $c_0$ is:

$$
\dot M \sim c_0^{3+\frac{2}{\gamma-1}}.
$$
If we transform the "energetic" overflow of the Roche lobe to "geometrical"
overflow using the formula

$$
c_0^2=\Delta\Phi=\frac{GM_1}{R_1^2}\Delta R,
$$
($R_1$ is the effective radius of the donor star; $\Delta R =
R_1 - R_{L_1}$, where $R_{L_1}$ is the radius of a sphere with
volume equal to the volume of the Roche lobe), then for
$\gamma = 5/3$, we obtain for the mass flow the widely-used
expression

$$
\dot M \sim (\Delta R)^3.
$$

   These expressions for the mass flow into the system were
derived in the framework of simplified models. Their common and,
obviously most serious, disadvantage is the lack of allowance
for the system's common envelope, which appreciably influences
the flow pattern along the Roche surface and, correspondingly,
the mass exchange parameters in the system.  This effect has
unfortunately previously been neglected, due to the fact that a
correct study of this phenomenon is possible only using
three-dimensional numerical gas dynamics models.

\section{Numerical Simulation Results}

   We base our analysis of the flow pattern near the donor star
filling its Roche lobe on the results described in Paper I. The
formulation of the problem is laid out in detail in Paper I, so
that we repeat here only the basic features of the model,
referring the reader to Paper I for details.

(1) We considered a low-mass X-ray binary similar to X1822--371
in which the mass of the outflowing component $M_1$ = 0.28
$M_\odot$, the temperature of the gas on the surface of this
component $T=10^4$K, the mass of the secondary (a compact object
with radius $0.05 R_\odot$) -- $M_2= 1.4 M_\odot$, the system
orbital period $P_{orb} = 1^d.78$ , and the distance between the
component centers $A = 7.35 R_\odot$.

(2) A three-dimensional system of gas dynamics equations closed
by the ideal gas equation of state was used to describe the
flow.

(3) To take into account radiative losses in the system, we took
the adiabatic index to be close to unity, namely $\gamma =
1.01$, which is close to the isothermal value [20].

(4) We assumed that the outflowing star fills its Roche lobe and
that the velocity of the gas at the surface is directed normal
to the surface. We took the gas velocity to be equal to the
local sound speed $u_0 = c_0 = 9$km/s. The density at the
surface of the outflowing component $\rho$ was taken to be
$\rho_0$.  Note that the boundary value of the density does not
affect the solution, due to the scaling of the system of
equations in $\rho$ and $P$. We used an arbitrary value for
$\rho_0$ in the calculations; when a specific system with a
known rate of mass loss is considered, the real densities in the
system can be determined simply by adjusting the calculated
density values using the scale indicated by the ratio of the
real and simulated densities at the surface of the outflowing
component.

(5) We adopted the conditions of free mass outflow both at the
accretor and at the outer boundary of the calculation region.

(6) The calculated region was the parallelepiped $(-A..2A)\times
(-A..A)\times(0..A)$; by virtue of the problem's symmetry
relative to the equatorial plane, the calculations were
performed only for the upper half-space.

(7) We used a high-order Total Variation Diminishing scheme to
solve the system of equations on a non-uniform (finer along the
line connecting the component centers) difference grid with $78
\times 69\times 35$ nodes.

The system of equations was solved starting from arbitrarily
chosen initial conditions up to the establishment of a
stationary flow regime. To ensure that this flow regime was
established, we continued the calculations over a rather long
time interval (exceeding ten orbital periods) after the
initiation of the stationary regime.

\begin{figure}
\plotone{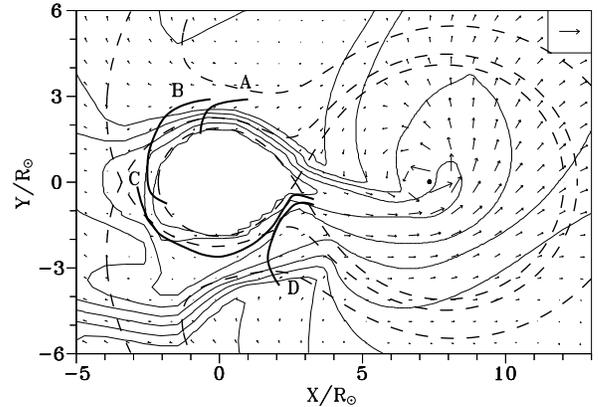}
\caption{Density isolines and velocity vectors in the equatorial
plane of the system. Roche equipotentials are shown with dashed
lines. Four stream lines marked with the letters A, B, C, D are
also shown, illustrating directions of the flow of the common
envelope material. The location of the accretor is marked with
the dark circle. The vector in the upper right corner
corresponds to a velocity of 800 km/s.}
\end{figure}

   The overall flow pattern in the equatorial plane of the
system considered is presented in Fig.~1, which depicts density
and velocity field contours in the area from -5 to 13$R_\odot$
along the $X$ axis and from -6 to 6$R_\odot$ along the $Y$ axis.
Figure 1 also shows four stream lines (marked with the letters
A, B, C, D), which illustrate the directions of flows in the
system.  Analysis of the results in Fig.~1 shows that a
significant fraction of the common envelope gas (stream lines A
and B) approaches the surface of the outflowing star (Roche
lobe) in the course of its motion, and is accreted by this
surface, preventing the formation of flows along the stellar
surface.  Some of the envelope material (stream lines C and D)
approaches to the donor star and, stripping gas off its surface,
participates in the formation of the stream in the vicinity of
$L_1$.

\begin{figure}
\plotone{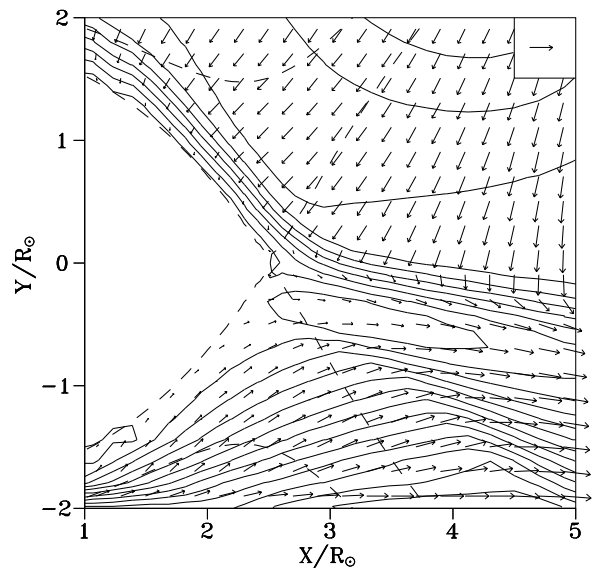}
\caption{Density isolines and velocity vectors in the equatorial
plane of the system in the vicinity of the inner Lagrange point.
Roche equipotentials are shown with dashed lines. The vector in
the upper right corner corresponds to a velocity of 300 km/s.}
\end{figure}

   Details of the flow pattern in the vicinity of the inner
Lagrange point are shown in Fig.~2, which shows the same flow
parameters as in Fig.~1 in a small region of the equatorial
plane from -1 to 5$R_\odot$ along the $X$ axis and from -2 to
2$R_\odot$ along the $Y$ axis. We can see from Fig.~2 that the
common envelope gas substantially changes the flow structure in
the vicinity of $L_1$, in particular, it strips matter off the
surface of the donor star. The asymmetry of the action of the
envelope gas on the outflowing stream is also visible in Fig.~2.
Gas flowing in from above, along the direction of the orbital
motion, is accreted by the donor star, and only in a small
region near $L_1$ is this matter taken up from the surface and
transported to the stream. The envelope gas, approaching to
$L_1$ from below, strips matter off a significant fraction
of the donor star surface.

  This effect of "stripping" matter off the surface of the donor
star substantially changes the generally accepted views of the
formation of the matter stream flowing into the system and of
the corresponding mass exchange parameters.  According to the
standard models, the structure of the atmosphere near the
surface of the donor star is determined by the equations of
hydrostatic equilibrium (see, e.g., [1, 18]).  For the adopted
system parameters and gas temperature (sound speed) at the donor
star surface, the gas energy is not large enough for the gas to
escape directly from the stellar surface, so that the mass flow
associated with thermal escape will be negligible small compared
to the flow through the vicinity of $L_1$. Gas located in the
near-surface layer (with characteristic height of the order of
the atmospheric scale height) can flow along the stellar
surface, but, in this case, also, the total mass flow into the
system does not change significantly [1].

   The situation fundamentally changes only when the system's
common envelope is taken into account, because, in this case,
gas from the surface layer can be "stripped" from the star and
carried into the system. Our calculations have shown that this
is the flow regime that is realized in the system considered
here. The momentum of the rarefied common envelope gas along the
surface of the star is large enough to pick up matter from the
surface layer over a significant fraction of the stellar
surface. Further, this material and the gas flowing out from the
vicinity of $L_1$ form the stream of matter flowing into the
system, and also determine the total mass flow.

\begin{figure}
\plotone{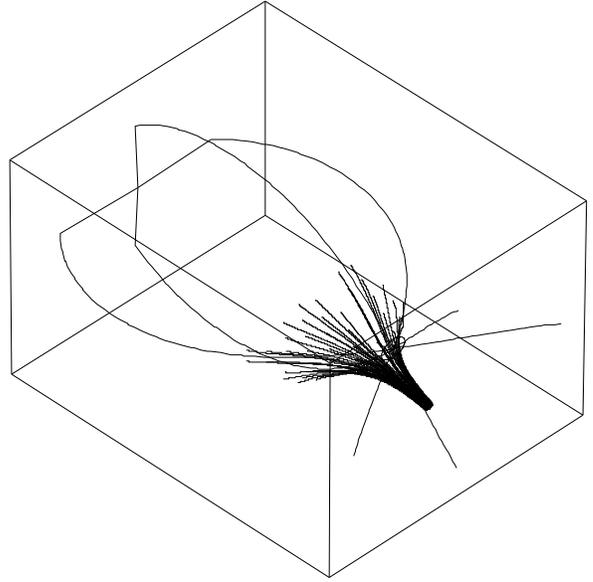}
\caption{Matter stream lines originating from the donor star
surface and forming the gas stream. The stream cross-section
obtained in standard models is shown by the solid ellipse near
$L_1$.}
\end{figure}

\begin{figure}
\plotone{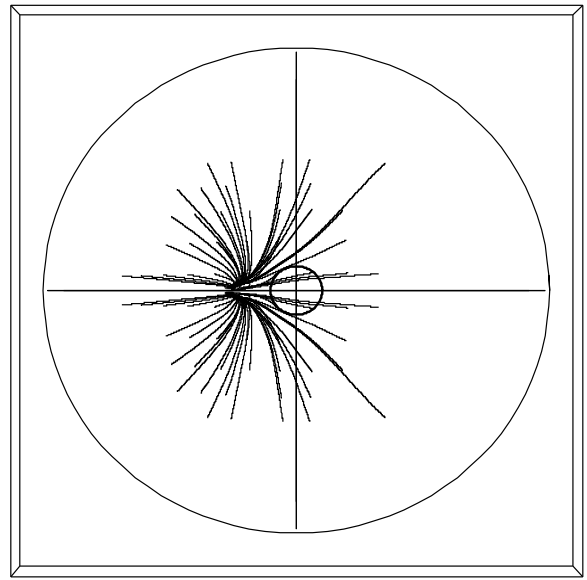}
\caption{View of the matter stream lines in Fig.3 for an
observer on the line connecting the centers of the two stars.
The stream cross-section obtained in standard models is shown by
the ellipse near $L_1$. The projection of the outflowing star on
the plane depicted is shown by the outer ellipse.}
\end{figure}

   The effect of the envelope gas on the flow structure near
$L_1$ is especially visible in Fig.~3, which shows a
three-dimensional projection of stream lines originating on the
stellar surface and entering the system with the gas stream. The
area containing the points through which the stream lines enter
the system represents that part of the surface providing mass
exchange in the system. We can clearly see the asymmetry of this
area in Fig.~4, which also shows stream lines leaving the
stellar surface (also in 3-D projection), but for an observer
located on the line joining the component centers. The size of
this area substantially exceeds the size of the stream in
standard models (shown by the ellipses near $L_1$ in Figs.~3
and 4). Analysis of these results indicates that the stream
forms in a region is more than order of magnitude larger than
predicted by theoretical estimates obtained without allowance
for the common envelope and with the same values for the gas
parameters at the donor star surface.
The matter flow injected into the system can now be represented
as the sum of two terms, and, accordingly, we obtain instead of
(4)

$$
\dot M ={\dot M}_{L_1} + {\dot M}_{Envel},\eqno(5)
$$
where the first term, derived from standard estimates, is an order
of magnitude smaller than the second term, determined by the
common envelope gas.

\figurenum{5a}
\begin{figure}[h]
\plotone{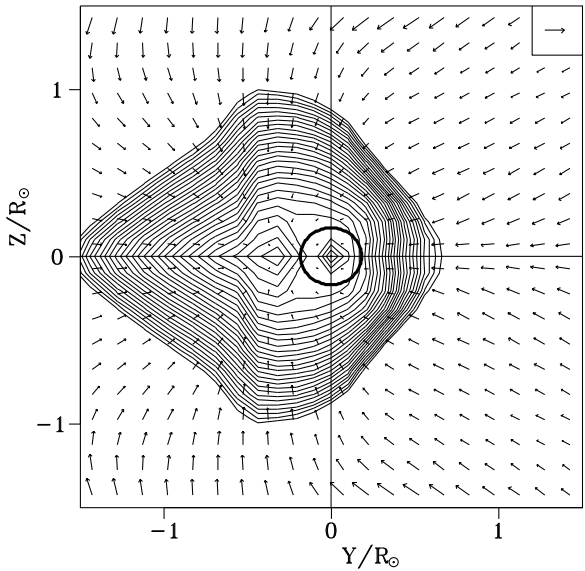}
\caption{Density isolines and velocity vectors in the stream
section formed by the $YZ$ plane passing through the inner
Lagrange point. The stream cross-section
obtained in standard models is shown by the bold ellipse.
 The vector in the upper right corner corresponds
to a velocity of 200 km/s. The maximum density (at the stream
center) is $1.2\rho_0$, and the minimum density (outer
contour) is $0.01\rho_0$.}
\end{figure}

\figurenum{5b}
\begin{figure}[h]
\plotone{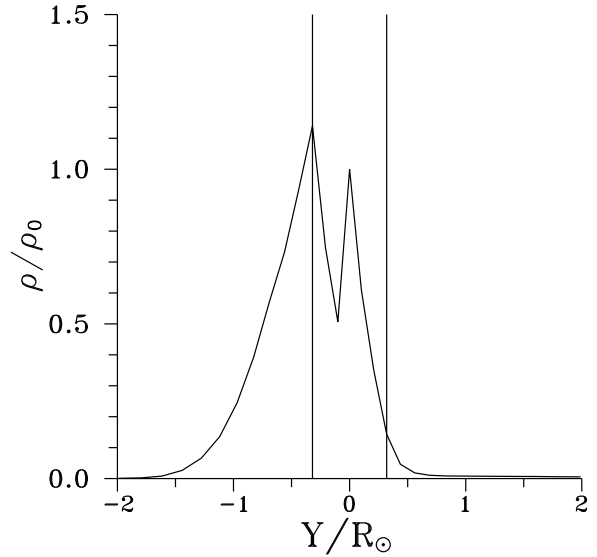}
\caption{Density distribution (normalized to the value at the
stellar surface) along the $Y$ axis for the stream cross section
presented in Fig.~5a. The stream boundaries obtained in standard
models are marked with vertical lines.}
\end{figure}

\figurenum{6}
\begin{figure}[h]
\plotone{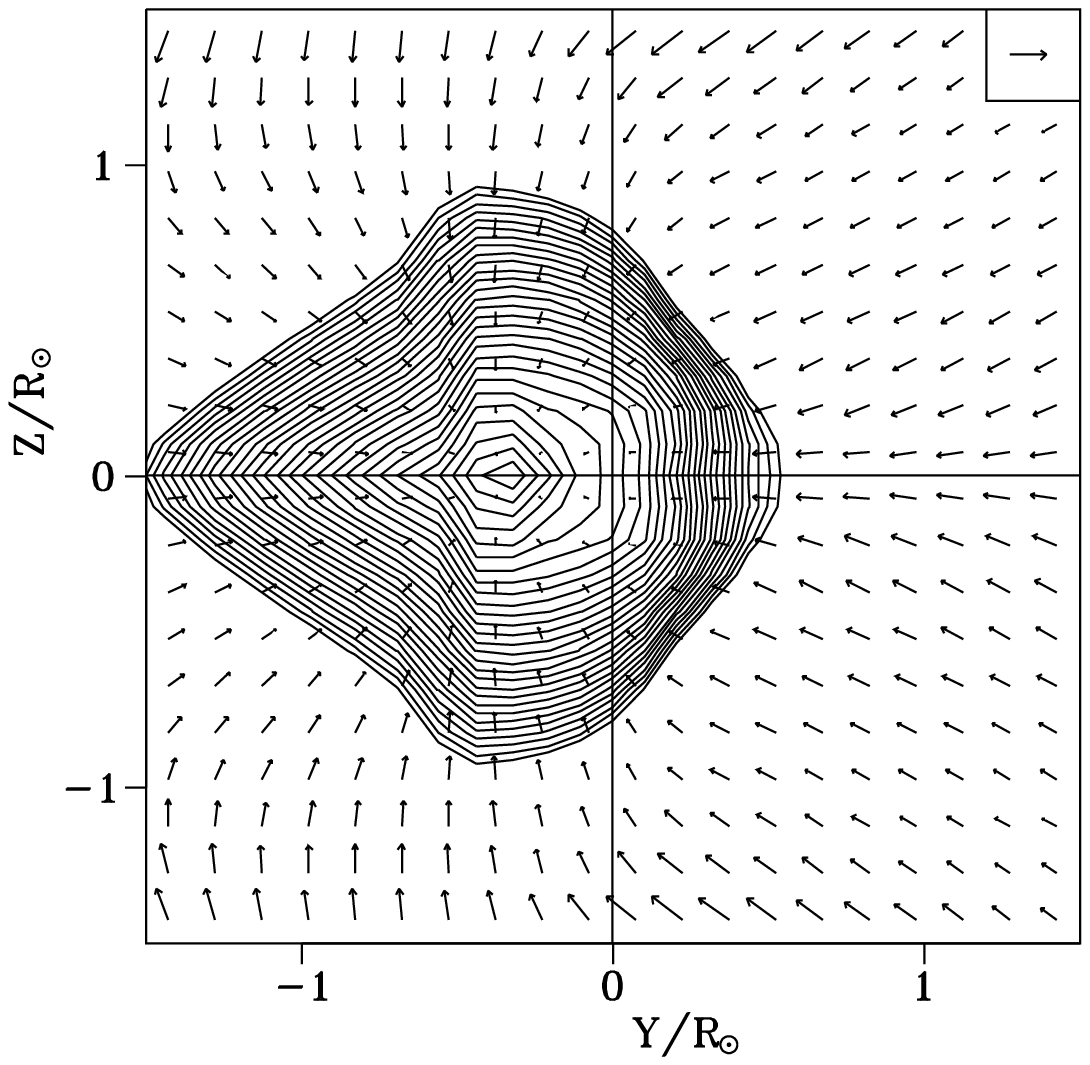}
\caption{Density isolines and velocity vectors in the stream
section formed by the $YZ$ plane passing $0.1R_\odot$ from the
inner Lagrange point. The vector in the upper right corner
corresponds to a velocity of 200 km/s. The maximum density (at
the stream center) is $1.4\rho_0$, and the minimum density
(outer contour) is $0.01\rho_0$.}
\end{figure}

 We will now consider the distributions of gas parameters
across the stream cross section. Figure 5 shows density and
velocity vector contours in the stream cross section formed by
the $YZ$ plane passing through $L_1$. The maximum density (in
the center) exceeds the density at the stellar surface $\rho_0$
by a factor of 1.2; the minimum density, depicted by the outer
contour, corresponds to $0.01\rho_0$. We can see from Fig.~5
that the matter stripped off the stellar surface by the common
envelope gas collects in a small zone centered on the equatorial
plane. The center of this zone, which is determined by the
common envelope gas flows considered above, does not coincide
with $L_1$ (the point $(0,0)$ in the figure); moreover, it lies
beyond the matter stream inferred from standard models (shown by
the ellipse near $L_1$ in Fig.~5). The two-humped nature of the
density distribution can be seen particularly well in Fig.~6,
which shows the density distribution (normalized to its value at
the stellar surface) along the $Y$ axis for the section of the
stream under consideration. The stream boundaries obtained in
standard models are shown in the figure by the vertical lines.
The effects described above can be interpreted as the presence
of two well-defined streams in the system, one associated with
the gas flowing from $L_1$, and the other forming under the
action of the common envelope gas.  Further analysis of the
calculation results indicates that the distributions of gas
parameters across the stream change with distance from the inner
Lagrange point, and the stream becomes a single flow at a
distance of the order of $0.1R_\odot$ ($0.013A$) from $L_1$.
This picture is supported by the results in Fig.~6, which shows
density contours in the $YZ$ cross section of the stream
$0.1R_\odot$ from the inner Lagrangian point. The
maximum density in Fig.~6 is $1.4\rho_0$, and the minimum
(outer contour) is $0.01\rho_0$.

\section{Conclusion}

   Analysis of the three-dimensional numerical simulation
results presented here indicates that taking into account the
impact of the common envelope in semidetached binary systems
radically changes the mass exchange parameters obtained. For the
low-mass X-ray binary X1822--371 that we have
considered, the overall mass inflow to the system increased by a
factor of about 50 compared to the estimates of standard models
for the same gas parameters values at the surface of the donor
star. In addition, the common envelope gas changed the flow
pattern near the surface of the outflowing component, which
ultimately affected the overall structure of the gas flows in
the system, and, thus, the expected observational manifestations
of these flows.

   Unfortunately, the quantitative estimates we have obtained
are valid only for the specific object considered, and do not
allow us to draw more general conclusions about changes in the
mass exchange parameters expected for other semidetached
systems.  This is due to the fact that the mass flow increase
depends not only on the parameters of the binary system, but
also on the properties of the common envelope gas, which can only
be determined using three-dimensional numerical simulations.
Nevertheless, given the results presented here, it is clear that
an adequate description of the mass flows in any semidetached
system with a common envelope is possible only in models that
take into consideration the influence of this envelope.

\acknowledgments

This work was supported by the Russian Foundation for Basic Research
(project code 96-02-16140).

\end{document}